\documentclass[preprint,12pt]{elsarticle}

\usepackage{amssymb}
\usepackage{subfigure}

\usepackage{lineno}
\usepackage{color}
\journal{NIM A}

\begin{document}

\begin{frontmatter}



\title{Charaterization of the first prototype IHEP-NDL LGAD sensor}
\author[label1]{Yuzhen Yang}

\author[label1,label3]{Suyu Xiao}
\author[label1]{Yunyun Fan}
\author[label4]{Dejun Han}
\author[label1,label2]{Zhijun Liang}
\author[label1]{Baohua Qi}
\author[label1]{Liaoshan Shi}
\author[label1,label3]{Yuhang Tan}
\author[label4]{Xingan Zhang}

\author[label1,label2]{Xin Shi\corref{cor1}}
\ead{shixin@ihep.ac.cn}

\address[label1]{Institute of High Energy Physics, Chinese Academy of Sciences, 19B Yuquan Road, Shijingshan District, Beijing 100049, China}
\address[label2]{State Key Laboratory of Particle Detection and Electronics, 19B Yuquan Road, Shijingshan District, Beijing 100049, China}
\address[label3]{University of Chinese Academy of Sciences, 19A Yuquan Road, Shijingshan District, Beijing 100049, China}
\address[label4]{Novel Device Laboratory, Beijing Normal University, No. 19, Xinjiekouwai Street, Haidian District, Beiing 100875, China}
\cortext[cor1]{Corresponding author}

\begin{abstract}
The timing measurement of charged particles using silicon detector is widely used in synchrotron source as X-ray detectors, in time-of-flight mass spectrometer and especially in large collider experiment. To reduce the drastically event pile-up of high-luminosity large hadron collider (HL-LHC), a new concept of 4-dimension detector including timing and positon has been proposed. One of the candidates for the 4-dimension detector is a new kind of silicon detector called Low Gain Avalanche Diode (LGAD). In China, Institute of High Energy Physics (IHEP) Chinese Academic Science cooperated with Novel Device Laboratory (NDL) at Beijing Normal University have fabricated a series of LGAD sensors. The characterization of the first prototype of IHEP-NDL sensors is presented including leakage current and sensor capacitance measurement. A test system for the time resolution of jitter term using pico-second laser and fast sampling rate oscilloscope is also setup, and the time resolution of 10 ps can be achieved with these sensors.

\end{abstract}



\begin{keyword}
Silicon detector \sep LGAD sensor \sep time resolution \sep pico-second laser


\end{keyword}

\end{frontmatter}


\section{Introduction}
\label{}
The timing measurement of charged particles using silicon detector is widely used in synchrotron source as X-ray detectors \cite{lgad_synchroton_application}, in time-of-flight mass spectrometer \cite{TOF_MS_silicon} and especially in large collider experiment. To reduce the drastically event pile-up of High Luminosity Large Hadron Collider (HL-LHC), a new concept of 4-dimension detector including timing and positon has been proposed \cite{4d_tracker}, such as High Granularity Timing Detector (HGTD) for ATLAS \cite{ATLAS_PhaseII}, Endcap Timing Layer (ETL) of MIP Timing Detector (MTD) for CMS \cite{MTD_CMS_TDR}. The best candidate for the 4-dimension detector is a new kind of silicon detector called Low Gain Avalanche Diode (LGAD) \cite{c_1st_LGAD_Pellegrinio_2014}\cite{d_LGAD_Cartiglia}. Similar technology would be also interested by the future Circular Electron Positron Collider (CEPC) for the particle indentification (PID) in flavor physics.

LGAD is a special Avalanche Photo Diodes (APD) with low gain (in the range of 10) to detect charged particles. The design is based on a modification of doping profile where an additional thin doping layer of p$^+$ material introduced close to the n-in-p junction, shown in Fig.\ref{fig:ihep-ndl-lgad}a. The large increase of doping concentration adjacent to the junction creates a large electrical field. The electrons drifting toward the n$^+$$^+$ electrode initiate the multiplication process, while the electric field is not enough for the avalanche mode so that the signal to noise is large \cite{doping_concentration}. In addition, to improve the time resolution, thin sensor of about 20$\sim$300$\mu$m thickness is designed for the fast rising edge of signal.

CERN RD50 collobration has initiated the LGAD research for future particle collider \cite{c_1st_LGAD_Pellegrinio_2014}\cite{d_LGAD_Cartiglia}. The LGAD technology was proposed and developed by the Centro Nacional de Microelectr\'{o}nica (CNM) Barcelona \cite{RD50_cnm}, Fondazione Bruno Kessler (FBK) and Hamamatsu Photonics (HPK) successively \cite{Hartmut_4D}. In China, Institute of High Energy Physics Chinese Academic Science (IHEP)cooperated with the Novel Device Laboratory (NDL) of Beijing Normal University also have fabricated a series of LGAD sensors.

In this paper,the basic characterization of the first prototype of IHEP-NDL sensors will be presented including leakage current measurement and capacitance measurement. We also setup a time resolution test system using pico-second laser and fast sampling rate oscilloscope.

\section{IHEP-NDL LGAD sensors: BV60 and BV170}

\begin{figure}
    \centering{
        \includegraphics[width=0.8\textwidth]{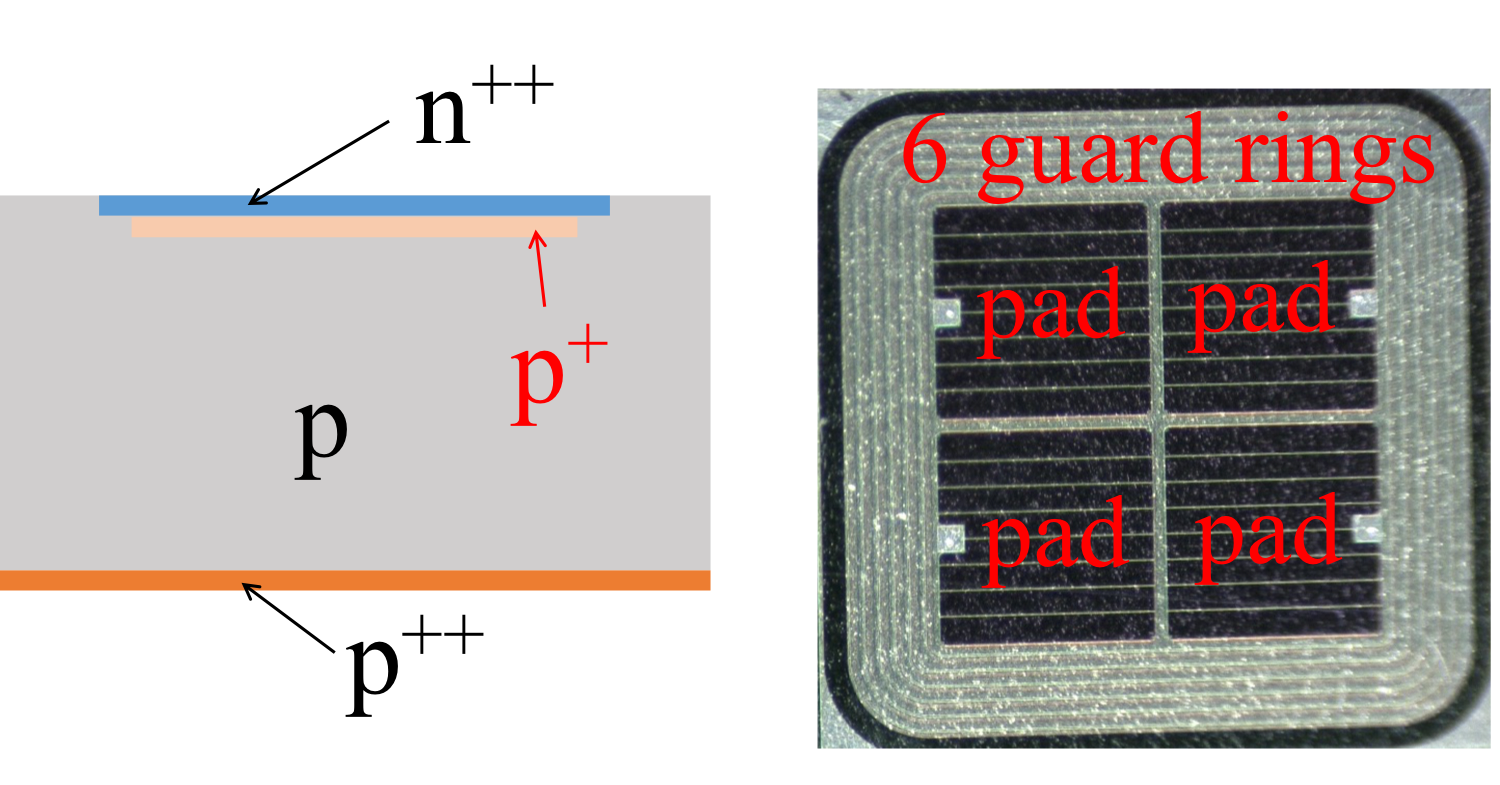}
        \put(-240,-10){\small(a)}
        \put(-80,-10){\small(b)}
        }
        \caption{IHEP-NDL LGAD sensor: (a) diagram of cross section structure, (b) picture of 2$\times$2 pads sensor.}
        \label{fig:ihep-ndl-lgad}
\end{figure}

The first prototype of IHEP-NDL LGAD sensor is n-in-p type with a additional p$^+$ layer close to the n-in-p junction, while the cross section view is shown in Fig.\ref{fig:ihep-ndl-lgad}a. With the same doping, there are two sets of IHEP-NDL LGAD sensors, denoted as BV60 and BV170 in the rest of this paper. Their volume resisitivity of silicon epitaxial layer are 300 and 100 $\Omega \cdot cm$ repectively. Both sensors are designed with 2$\times$2 pads and 6 floating guard rings, shown in Fig.\ref{fig:ihep-ndl-lgad}b. The pad size is about 1$\times$1 mm. Several aluminium lines on the surface of each pads are designed for the uniformity of bias voltage applied on the sensor. The thickness of epitaxial layer is 33 $\mu$m.

\section{Property of IHEP-NDL LGAD sensor}

\subsection{Leadage current and gain}
\begin{figure}
    \centering{
        \includegraphics[width=0.48\textwidth]{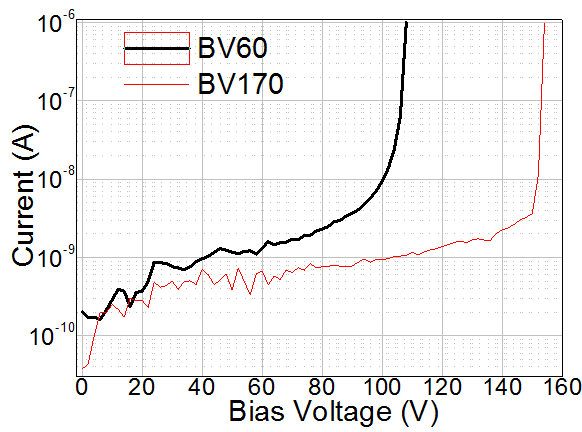}
        \put(-100,-10){\small(a)}
        \includegraphics[width=0.48\textwidth]{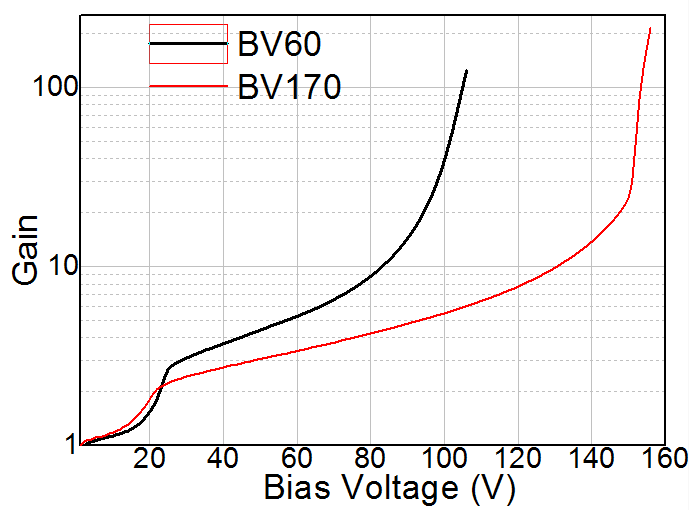}
        \put(-90,-10){\small(b)}
        }
    \caption{Measurement results of BV60 and BV170: (a)dark current vs. bias voltage, (b) gain vs. bias voltage.}
    \label{fig:iv_gain}
\end{figure}

The leakage current of LGAD sensor depends on the minority current carriers, and determine the working condition and power consumption of the detector curcuit. The current was measured by a source-meter on the metal isolated chuck of probe station. The relationship of leakage current to bias voltage tested at room temperature is shown in Fig.\ref{fig:iv_gain}a. The breakdown voltages of BV60 and BV170 are about 95 V and 160 V, respectively.

The gain of LGAD sensors can be measured by determining both the initial number of particles and the total number of particles after multiplication \cite{Hartmut_4D}. Generally, the gain is calculated by comparing to an identical sensor without the multiplication layer or to an known energy loss of minimum ionizing particles (MIPs). However, we assume that the LGAD sensor is without gain at 1 V bias voltage, and the initial current is the current difference between laser on and off while the bias voltage is 1 V. The rough gain was calculated through deviding the ouput current with laser on to the initial current. The multiplication process and gain of LGAD sensor are affected by the electrical field of multiplication layer, which depends on the bias voltage and doping profile. The rough gain of BV60 and BV170 as functions of bias voltage are shown in Fig.\ref{fig:iv_gain}b. The dependence of gain and leakage current on bias voltage are both almost exponential, especially when the gain is 3 $\sim$ 10, since the drift of charges generated by input particles and leakage current is similar. 

\subsection{Capacitance and doping profile}
\begin{figure}
    \centering{
    \includegraphics[width=0.48\textwidth]{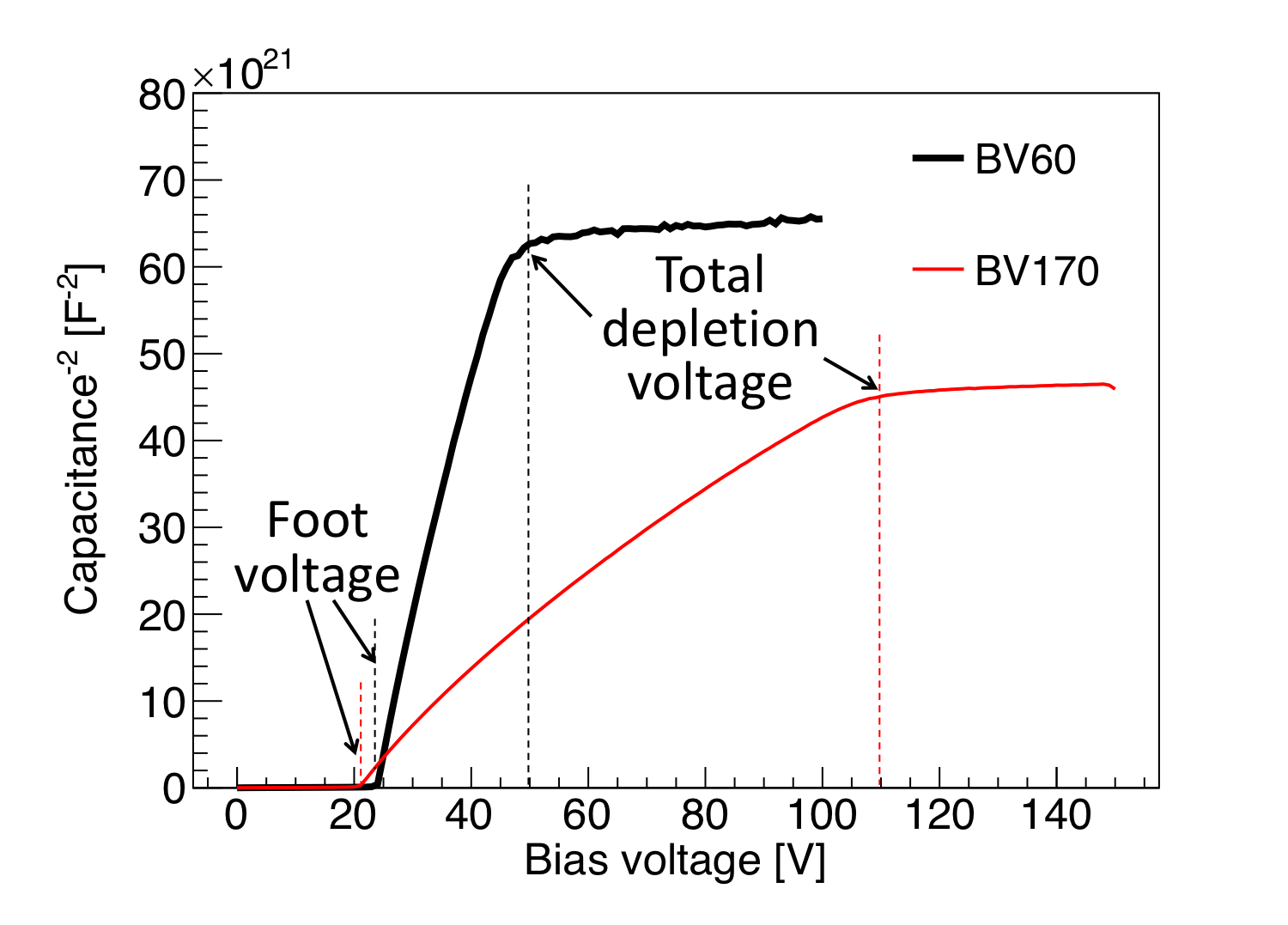}
    \put(-70,-10){\small(a)}
    \includegraphics[width=0.48\textwidth]{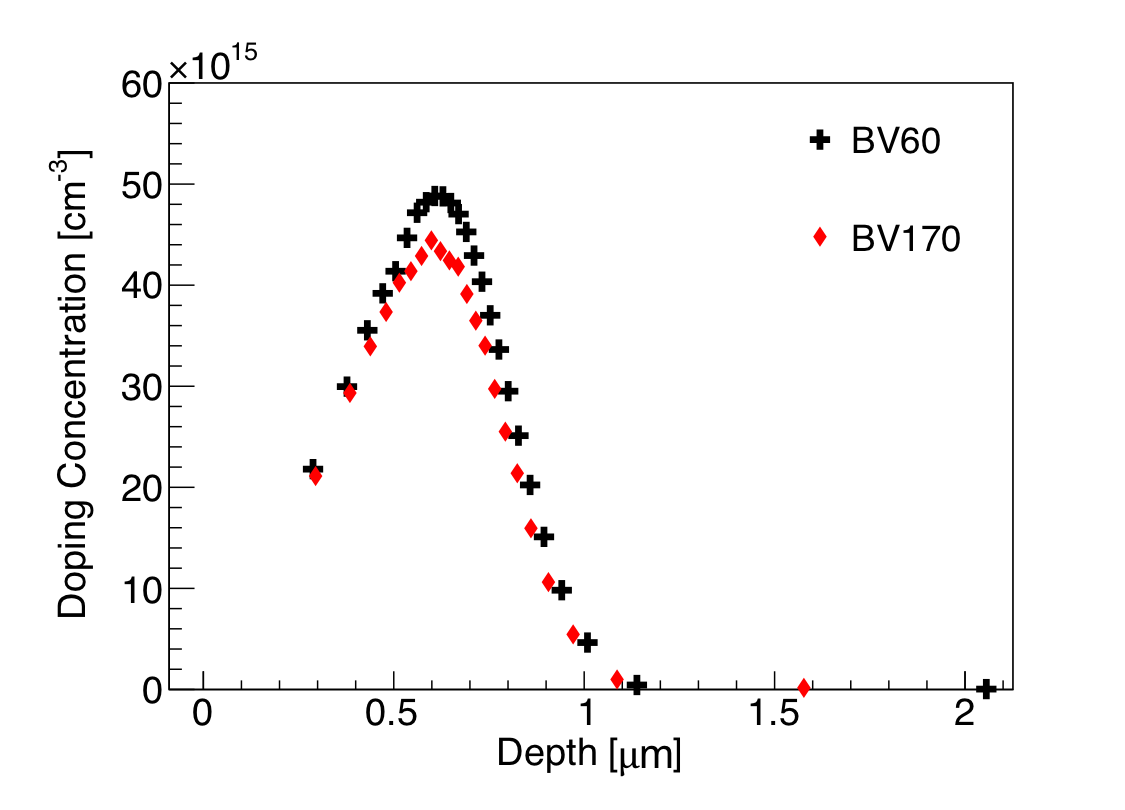}
    \put(-70,-10){\small(b)}
    }
    \caption{(a) 1/$C^{2}$ vs. bias voltage and (b) doping concentration vs. depth for BV60 and BV170.}
        \label{fig:doping}
\end{figure}

The capacitances $C$ of BV60 and BV170 with different bias voltages are measured by a precise LCR meter with a power supply on the isolated chuck of probe station. The measurements are performed at room temperature with a frequency of 10 kHz. The relationship of 1/$C^{2}$ to the applied bias voltage are shown in Fig.\ref{fig:doping}a. The foot voltage is the bias voltage needed to completely depleted the multiplication layer, which is related to the doping concentration of the multiplication layer. The foot voltage of BV60 and BV170 are about 24 V and 21 V, respectively. The capacitance decreases with the increase of reverse voltage, until the epitaxial layer is fully depleted. The total depletion voltage of BV60 and BV170 are 50 V and 110 V, respectively. Then, the capacitance becomes a constant due to little contribution of the spread of depletion region.

The doping profile of LGAD structure can be obtained from the capacitance $C$ measured as a function of bias voltage. The depth of multiplication layer and the capacitance variation with the applied bias voltage depend on doping concentration. The doping concentration $N(w)$ is calculated from the following equation \cite{doping_profile}:
\begin{eqnarray}
N(w)=-\frac2{q\epsilon_r \epsilon_0 A^2}[\frac{d}{dV}(\frac1{C^2})]^{-1}
\end{eqnarray}
with
\begin{eqnarray}
w=\epsilon_r \epsilon_0 \frac{A}{C}
\end{eqnarray}
where $q$ is the electron charge, $\epsilon_0$ is the vacuum absolute permittivity (8.854 $\times$10$^{-14}$ F/cm), $\epsilon_r$ is the silicon relative permittivity (equals 11.7), $A$ is the area of the pad (1$\times$1 mm$^{2}$). The doping profile is plotted as functions of depth in Fig.\ref{fig:doping}b. The depth of multiplication layer for BV60 and BV170 are about 0.6 $\mu$m.

\section{Time resolution}

\subsection{Time-tagging detector}
The time-tagging detectors can be used a simplified model to describe the timing capabilities \cite{Hartmut_4D}. The sensor can be treated as a capacitor with a current source in parallel, and is readout by a preamplifier to shape the signal. Then, the output of preamplifier is compared to a threshold to determine the arrival time. The time resolution $\sigma_t$ can be expressed as a sum of the following terms:
\begin{eqnarray}
\sigma_t^2 = \sigma_{Landau~noise}^2 + \sigma_{Signal~distortion}^2 + \sigma_{Time~walk}^2 + \sigma_{TDC}^2 + \sigma_{Jitter}^2
\end{eqnarray}
where $\sigma_{Landau~noise}$ is caused by the non-uniform energy deposition created by the ionizing particle passing a sensor. This variation also produces an overall change of signal magnitude.

$\sigma_{Signal~distortion}$ represents the uncertainty caused by the non unifrom drift velocity and weighting field according to the Ramo-Shockley's theorem \cite{Hartmut_4D}. It states the current induced by a charge carrier is proportional to its electric charge, the drift velocity and weighting field. To obtain uniform drift velocity of carriers in the volume in sensor, the easiest method is to produce an electric field high enough for the carriers to move with saturated velocity. On the other hand, to reduce the weighting field difference caused by the electrode geometry, eletrode needs to have a size very similar to the pitch and much larger than the sensor thickness.

$\sigma_{Time~walk}$ is caused by the earlier arrival time of larger signals crossing a given threshold than smaller signals. This effect can be compensated by an appropriate readout electronic circuit, such as Constant Fraction Discriminator (CFD), Time over Threshold (ToT) and Multiple Samplings (MS) \cite{the_4d_pixel_challenge}.

$\sigma_{TDC}$ is the contribution of TDC binning, which is usually below 10 ps and therefore it will be neglected \cite{Hartmut_4D}.

$\sigma_{Jitter}$ is caused by the early or late firing of the comparator due to the presence of noise on the signal or in the electronics. It is proportional to the system noise N and inversely proportional to the slope of signal around the threshold. Assuming a constant slope, we can write dV/dt $\approx$ S/t$_{rise}$ (where S is the amplitude of signal, t$_{rise}$ is the rise rime) and therefore $\sigma_{Jitter}$=N/(dV/dt) $\approx$ t$_{rise}$/(S/N) \cite{Hartmut_4D}. So the jitter term depends on the noise and the rise time of signal. We will focus on the effect of jitter term using the laser test in this paper.

\subsection{Experiment setup}

\begin{figure}
    \centering{
        \includegraphics[width=0.6\textwidth]{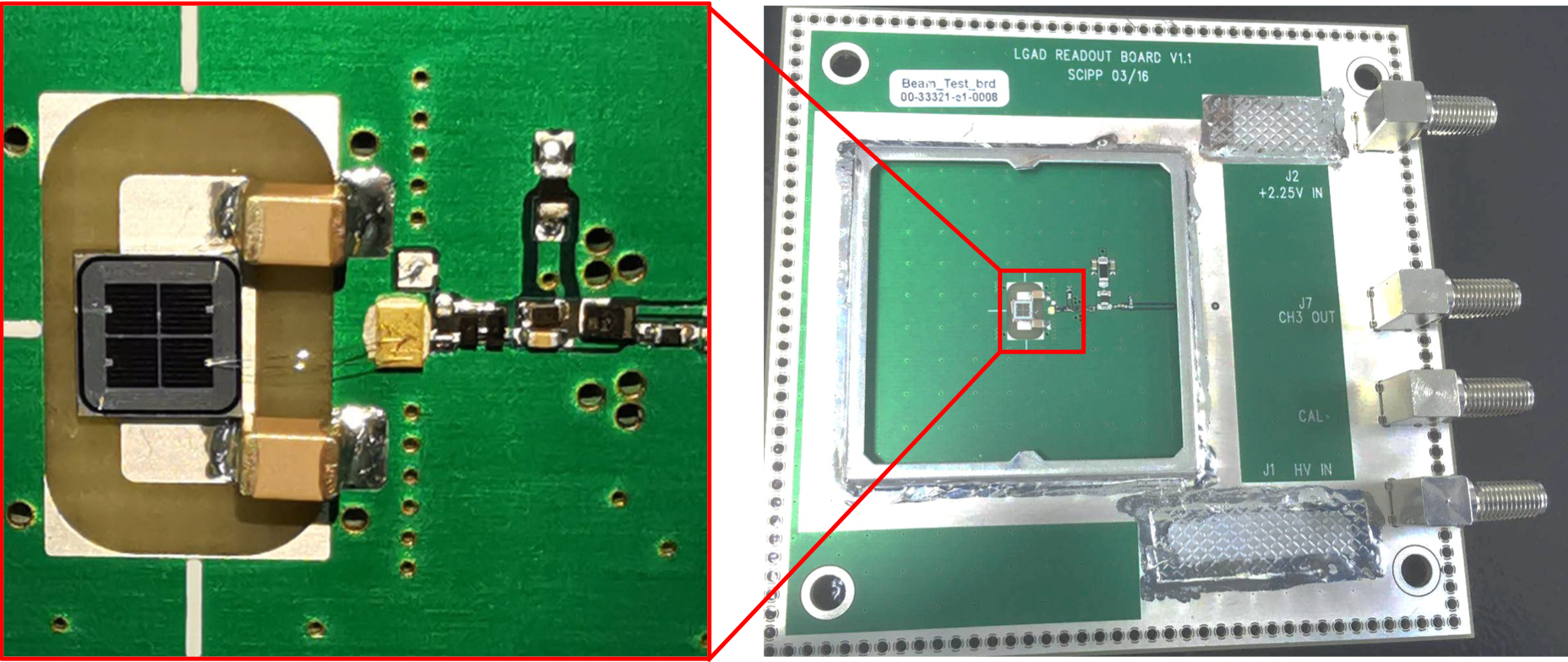}
        \put(-190,-10){\small(a)}
        \put(-70,-10){\small(b)}
        }
        \caption{Picture of IHEP-NDL LGAD sensor on readout board: (a) bonding wire connecting sensor to readout board, (b) overview.}
        \label{fig:readout}
\end{figure}

To study the time resolution, BV60 and BV170 were mounted on a 10$\times$10 cm$^{2}$ single channel read-out board developed at the University of California Santa Cruz, shown in Fig.\ref{fig:readout}b. The read-out board allows maintaining a wide bandwidth of $\sim$2 GHz and a low noise. It has been used for the test beam and the detail description can be found in \cite{16ps_testBoard,neutron_irradiation}. The sensor with 2$\times$2 pads is attached on the read-out board using double-side conductive tape. The front side metallization layer of one pad is coupled to the input of amplifier, through several aluminium bonding wires to minimize inductance, shown in Fig.\ref{fig:readout}a. The signals are recorded by a oscilloscope with a sampling rate of 40 GS/s and a bandwidth of 2.5 GHz. The vertical scale of scope affects the digitization noise which contributes to the overall noise, so that a relative large vertical scale is used for large signals at high gain.

\begin{figure}
    \centering{
        \includegraphics[width=0.52\textwidth]{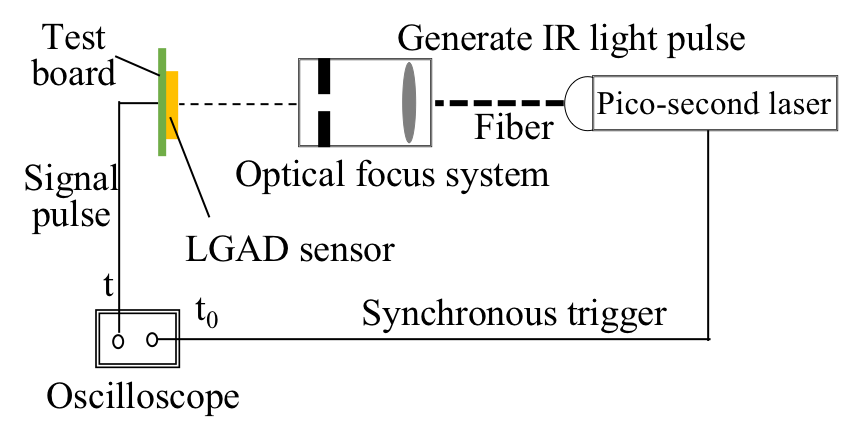}
        \put(-100,-10){\small(a)}
        \includegraphics[width=0.47\textwidth]{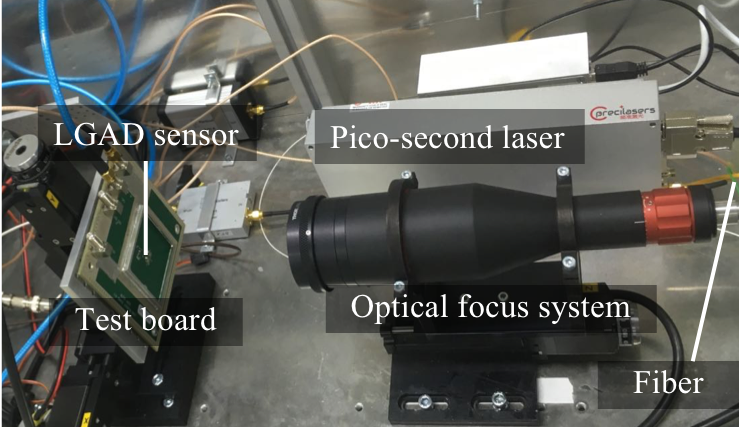}
        \put(-100,-10){\small(b)}
        }
        \caption{Time resolution test system using pico-second laser: (a) diagram, (b)picture.}
        \label{fig:laser}
\end{figure}

The diagram and photo of the test system for time resolution is shown in Fig.\ref{fig:laser}, using a pico-second laser with a wavelength of 1064 nm, a minimum light pulse width of 7.5 ps and a frequency of 20 MHz. A series of extremely short inferred (IR) light pulses are produced by the pico-second laser, focused and attenuated by the optical system, then illuminated on the front of LGAD sensor. Electron-hole pairs are produced and multiplied in the sensor, drifting and diffusing separetely, then collected by electrodes and multiplied by the amplifier on read-out board. The electrical signal is recorded by the oscilloscope using the trigger of synchronous pulses produced by the pico-second laser.

\subsection{Output signal}
The output signals of sensor and triggers in each event were recorded by the digital oscilloscope, so the complete event information is available through offline analysis.

\begin{figure}
    \centering{
        \includegraphics[width=0.49\textwidth]{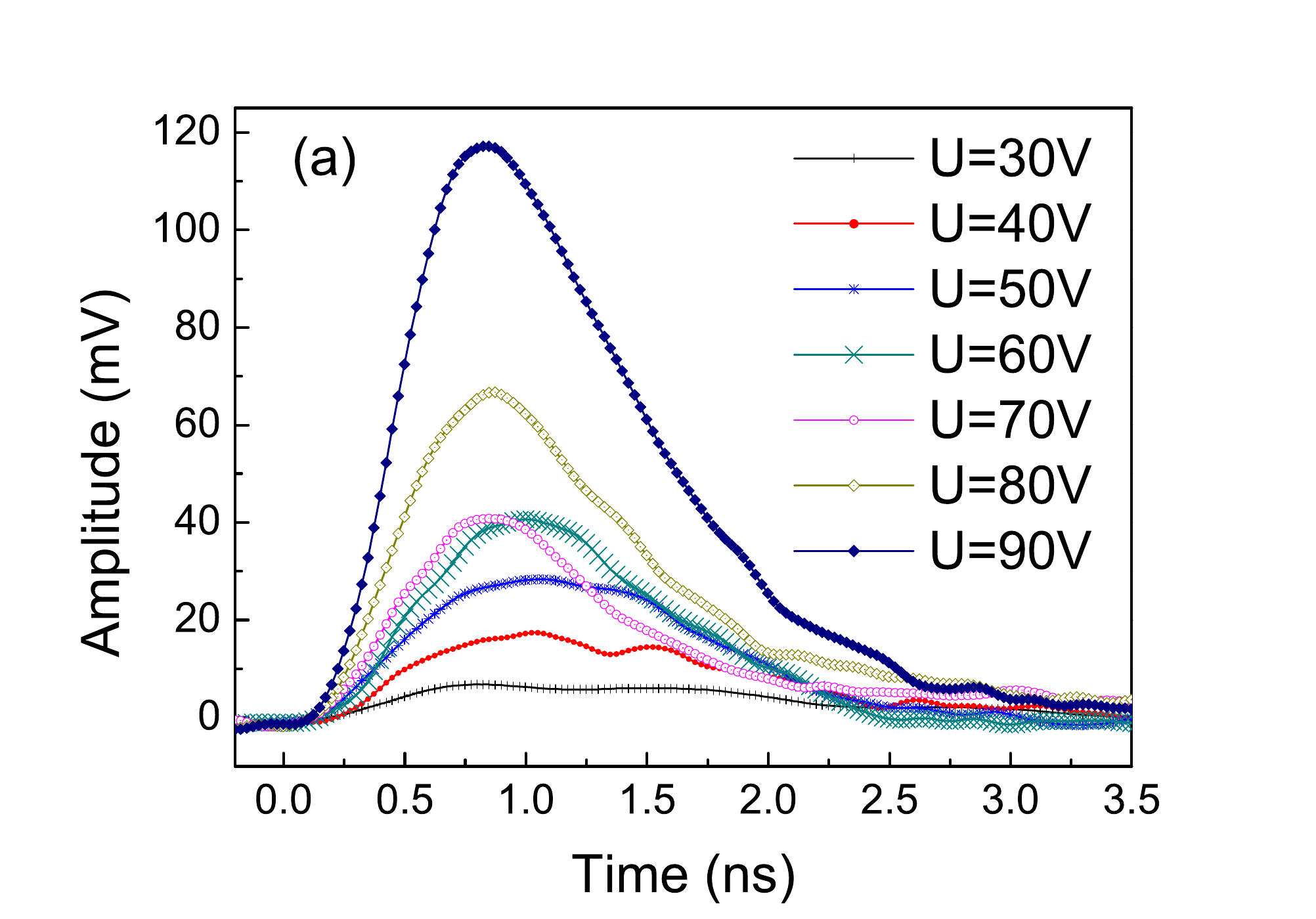}
        \includegraphics[width=0.49\textwidth]{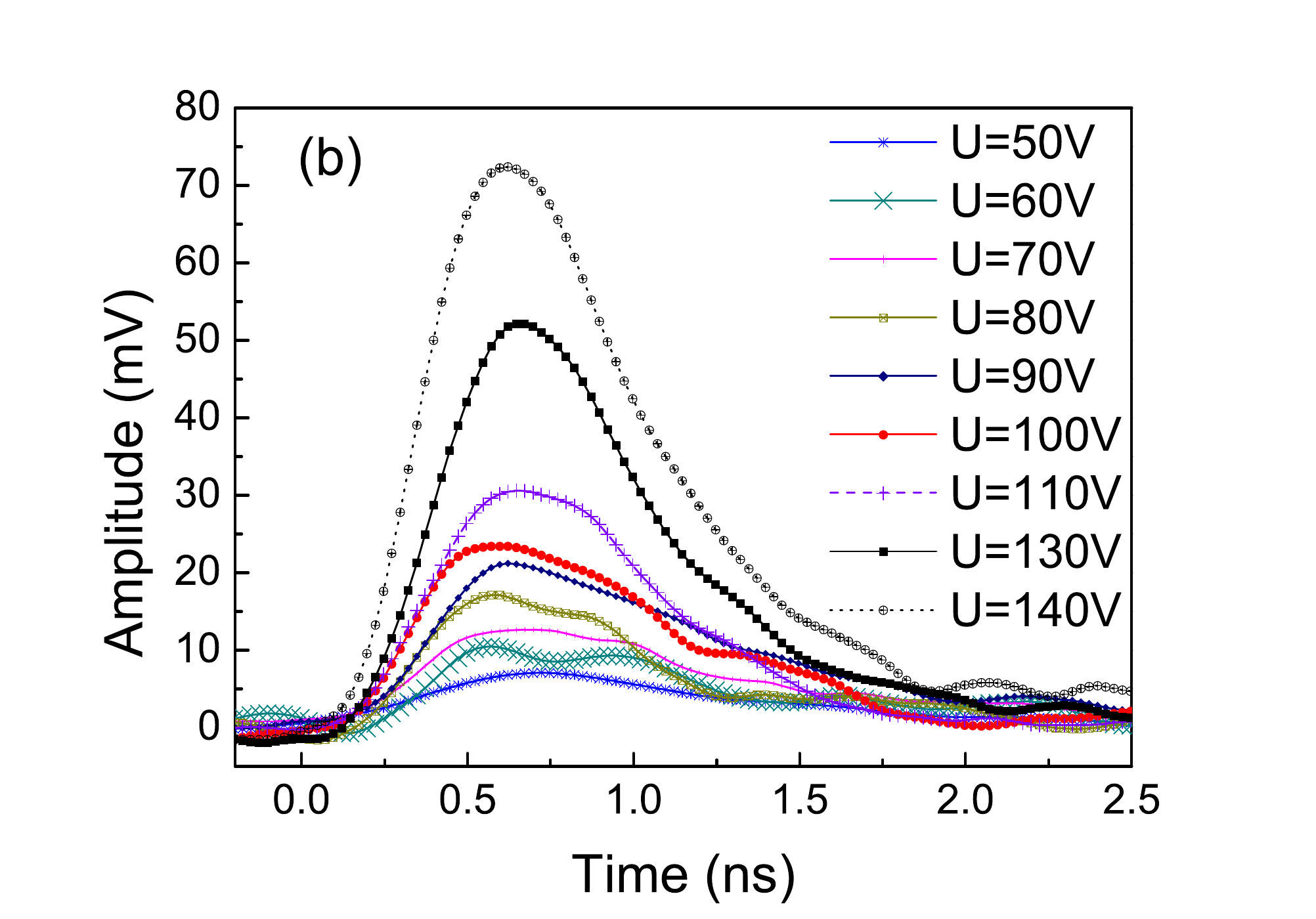}
        }
        \caption{Output signals of sensor with different bias voltages: (a) BV60, (b) BV170.}
        \label{fig:amp-time}
\end{figure}

The output signals of BV60 and BV170 after the amplifier on read-out board, with different bias voltages are shown in Fig.\ref{fig:amp-time}a and b, respectively. With the increase of bias voltage, the amplitudes of both sensors increase due to the higher gain at higher bias voltage; but the rise time decreases due to the increase of drift velocity. The tail of output signal is obvious at low bias voltage, since the bias voltage is too low for the saturation of the carriers drift velocity, while the LGAD sensor is not fully depleted as shown in Fig.\ref{fig:doping}b.

\subsection{Analysis method}
\begin{figure}
    \centering{
        \includegraphics[width=0.49\textwidth]{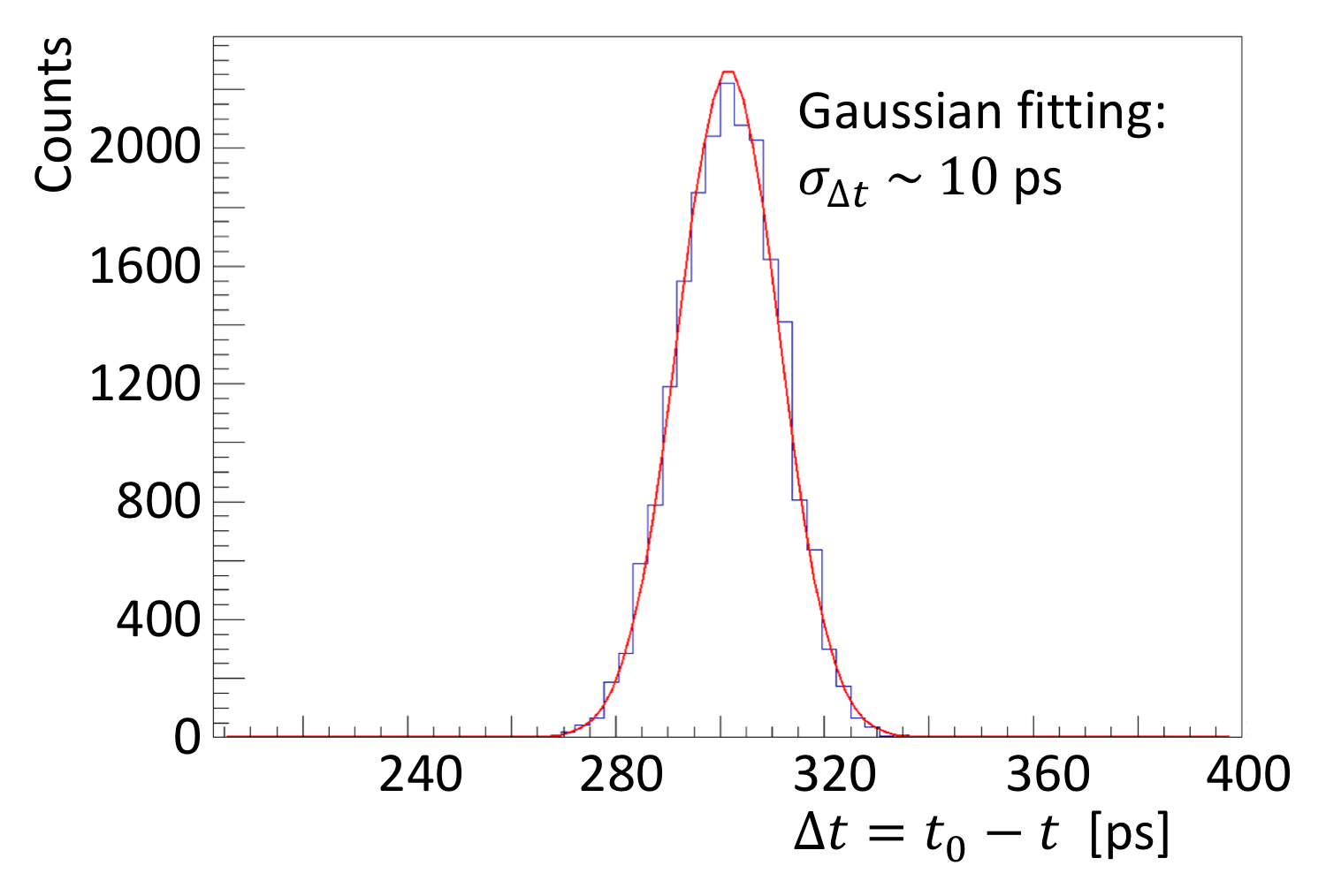}
        }
    \caption{Gaussian function fit for $\Delta t$ distribution of BV170 
    with 130V bias voltage to caculate the time resolution.}
    \label{fig:delta_t_fit}
\end{figure}

To reduce the time walk, CFD is a common method which sets the arrival time as the signal reaches a given fraction of the total amplitue. To optimize the time resolution, the constant fraction (CFD value) of trigger and signal amplitutes were studied with different bias voltages due to the contribution of pulse shape and noise change. Since the data of output signals is digital, the arrival time with specific CFD value is evaluated with a polynomial fitting function. The time difference between trigger and sensor signal is $\Delta t=t_0 -t$, where $t_0$ and $t$ are the arrival time of trigger and signal, respectively. The distribution of $\Delta t$ is fitted by a Gaussian function, and $\sigma$ of the Gaussian is defined as time resolution of jitter term. The $\Delta t$ distribution of BV170 sensor at 130V bias voltage is shown in Fig.\ref{fig:delta_t_fit}. The time resolution of jitter term is about 10 ps.

\subsection{Jitter term for LGAD time resolution}

\begin{figure}
    \centering{
    \includegraphics[width=0.49\textwidth]{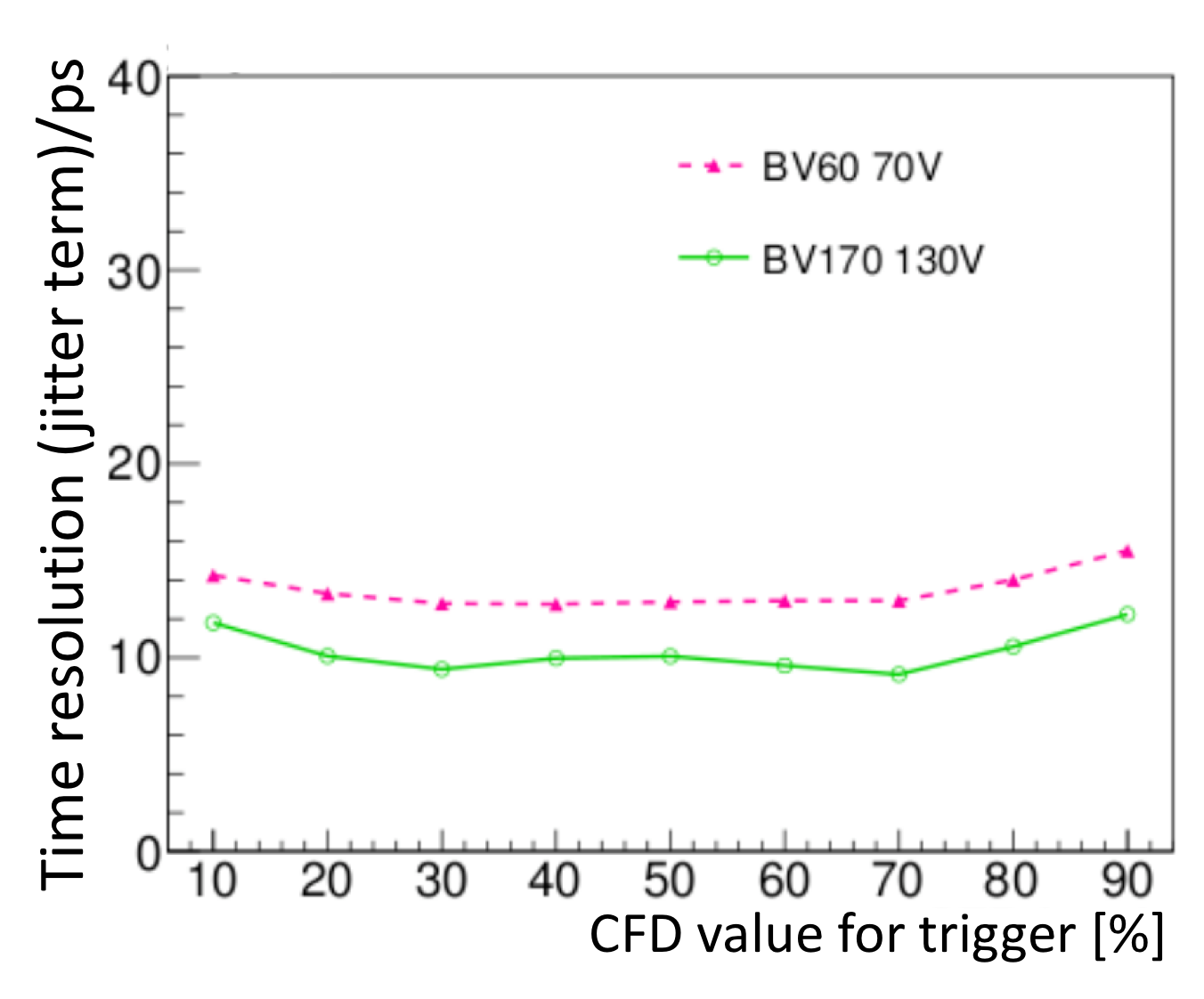}
    \includegraphics[width=0.49\textwidth]{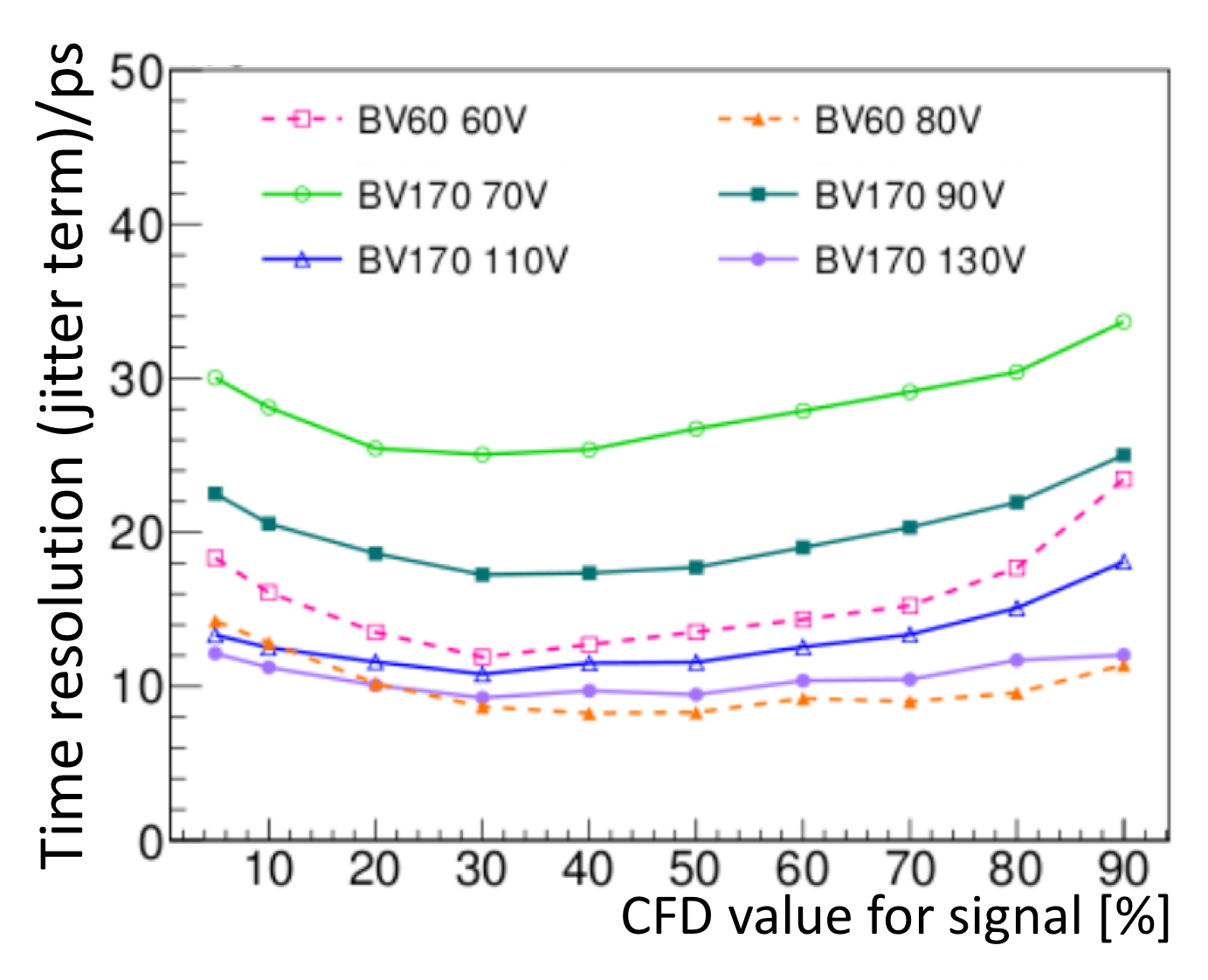}
        }
    \caption{Time resolution of jitter term as functions of (a) trigger and (b) signal CFD value with different bias voltage on BV60 and BV170.}
    \label{fig:time_res}
\end{figure}

For the measurement of time resolution of LGAD sensor using laser test, the number of input photon is stable and large enough to neglect the effect of Landau noise, and the time walk is mitigated by the CFD method, so jitter term plays the main role for time resolution. The effect of CFD values on jitter term of IHEP-NDL sensor with different bias voltages has been studied. The CFD value of trigger affects little on the jitter of BV60 and BV170 shown in Fig.\ref{fig:time_res}a, since the synchronous trigger generated by pico-second laser is stable enough.

The jitter of BV60 and BV170 with different bias voltages as a function of signal CFD values is shown in Fig.\ref{fig:time_res}b, while the trigger CFD value is set as 30\%. The jitter is inversely proportional to the slope of signal, so that the jitter contribution increases both at low and high CFD values while the signal shape is less steep. The trend is consistent with the simulation result in Reference \cite{Hartmut_4D}. It is reasonabe to set the signal CFD value as 30\%$\sim$70\%, while the slope of signal is relatively steep. In addition, the higher bias voltages produce the lower jitter due to the higher gain and amplitude of output signal.

\begin{figure}
    \centering{
        \includegraphics[width=0.6\textwidth]{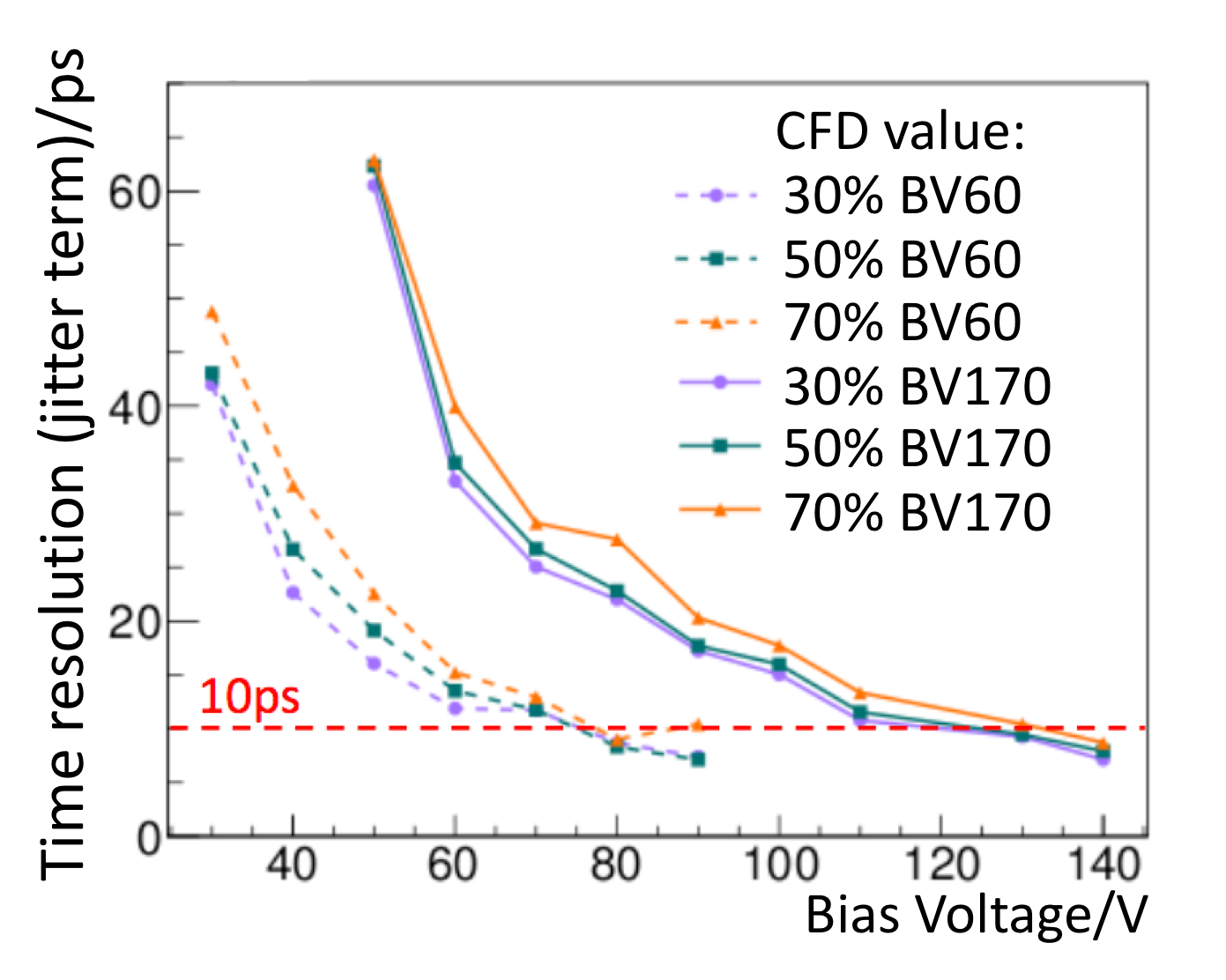}
        }
    \caption{Time resolution of jitter term as a function of bias voltage for BV60 and BV170.}
    \label{fig:time_res_bias}
\end{figure}

The jitter contribution of BV60 and BV170 as functions of bias voltages with CFD values of 30\%, 50\% and 70\% is shown in Fig.\ref{fig:time_res_bias}, while the trigger CFD value is set as 50\%. With the increase of bias voltage, jitters of both sensors become better, since the electrical field in sensor volume is stronger and the gain is higher. Also, the velocity of carriers becomes saturated with higher bias, and contribution of distortion to the time resolution becomes less. The jitter of both LGAD sensors achieve a saturation value, which is smaller than 10 ps at high bias voltage.

\section{Conclusion}
The performance of IHEP-NDL LGAD sensor BV60 and BV170 has been measured, including leakage current, gain and capacitance as the function of bias voltage. The test system for time resolution of jitter term is setup using pico-second laser and fast sampling rate oscilloscope. The jitter term of time resolution has achieved lower than 10 ps.

The Landau fluctuation is another significant term for time resolution especially for the particle detection. The time resolution study using beta source and test beam will also be performed in the near future.

In addition, the irradiation hardness of LGAD sensor play a key role for the detector life. So the performance of sensor affected on the proton flunce and X-ray dose will also be studied in the future.





\section*{Acknowledgement}
This work is supported by the State Key Laboratory of Particle Detection and Electronics in IHEP. We acknowlege the technical help by the CERN RD50 collaboration and ATLAS collaboration.

\section*{References}


\end{document}